\documentclass[12pt]{article}
\usepackage{epsf}
\usepackage{graphicx}

\begin{document}
\title
{Another possible interplay between gravitation and cosmology}
\author
{Michael A. Ivanov \\
Physics Dept.,\\
Belarus State University of Informatics and Radioelectronics, \\
6 P. Brovka Street,  BY 220027, Minsk, Republic of Belarus.\\
E-mail: michai@mail.by.}

\maketitle

\begin{abstract}
I describe here some features of a non-geometrical approach to
quantum gravity which leads to another picture of ties of
gravitation and cosmology. The role of taking into account the
effect of time dilation of the standard cosmological model is
considered. It is shown that the correction for no time dilation
leads to a good accordance of Supernovae 1a data and predictions
of the considered model. The distributions of stretch factor
values of Supernovae 1a for the cases of time dilation and no time
dilation are discussed.
\end{abstract}
The general theory of relativity and the standard cosmological
model of our time are connected very closely via the main idea of
a cosmological expansion. Their interplay engenders such strange
and "dark" concepts as Big Bang, inflation, dark energy and dark
matter. The last of such fantoms is dark flow \cite{771}; the
authors try to interpret in a frame of the standard model the
observed motion of galaxy clusters as a result of interaction with
another bubble of a multiverse (it is necessary to have a very
hard belief in the current paradigm to introduce such the
explanation as the first one). Of course, it is difficult to find
some other explanation of observed flat rotation curves of
galaxies and related phenomena than dark matter, but, perhaps, it
is not impossible. But in the case of inflation and dark energy,
the ones are obvious buttresses of the standard model in the
troubles.
\par
There is a very small, but iron made, effect which frustrates the
harmony of this connection: the Pioneer anomaly \cite{1}. It is
impossible to embed the one in a frame of general relativity; from
another side, a magnitude and a sign of this effect (the probe's
deceleration is approximately equal to $ Hc$, where $c$ is the
light velocity and $H$ is the Hubble constant) overshade seeming
successes of the current cosmological model.
\par
\begin{figure}[th]
\epsfxsize=13.0cm \centerline{\epsfbox{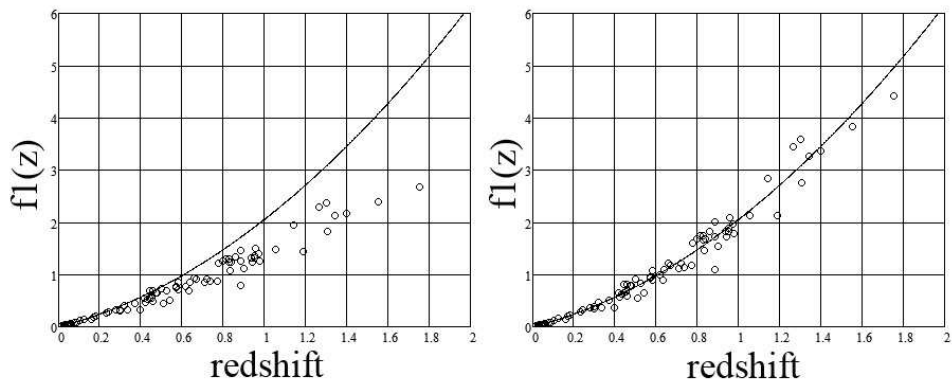}} \caption{
The theoretical function $f_{1}(z)$ of this model with $b=2.137$
(solid); Supernovae 1a observational data from Table 5 of
\cite{203} transformed to the linear scale (circles, 82 points):
corrected for time dilation (left panel) and corrected for no time
dilation (right panel).}
\end{figure}
I describe here some features of a non-geometrical approach to
quantum gravity \cite{500} which leads to another interplay of
gravitation and cosmology. My model is based on the idea of an
existence of the background of super-strong interacting gravitons.
An interaction of light with this background gives a specific
redshift mechanism which does not need any cosmological expansion;
its peculiarity is an additional relaxation of any light flux that
may be connected with the observed deviation of the Hubble diagram
from its expected view without dark energy in the standard model.
Due to this relaxation, any observer can see only a part of the
universe; the property is sufficient to explain the very important
results of observations of a bulk flow of clusters reported in
\cite{771} without any exotic and dark names. In the model, the
Newton and Hubble constants may be computed. An important feature
of the model is an essential difference of inertial and
gravitational masses of black holes; it means that an existence of
black holes contradicts to the equivalence principle.
Additionally, the property of asymptotic freedom of this model at
very short distances leads to the important consequence: a black
hole mass threshold should exist \cite{216,217}. A full mass of
black hole should be restricted from the bottom with $m_{0}$; the
rough estimate for it is: $m_{0} \sim 10^{7}M_{\odot}$. The range
of transition to gravitational asymptotic freedom for a pair of
protons is between $10^{-11} - 10^{-13}$ meter, while for a pair
of electrons it is between $10^{-13} - 10^{-15}$ meter. This
transition is non-universal \cite{216}; it means that a
geometrical description of gravity on this or smaller scales, for
\begin{figure}[th]
\epsfxsize=13.0cm \centerline{\epsfbox{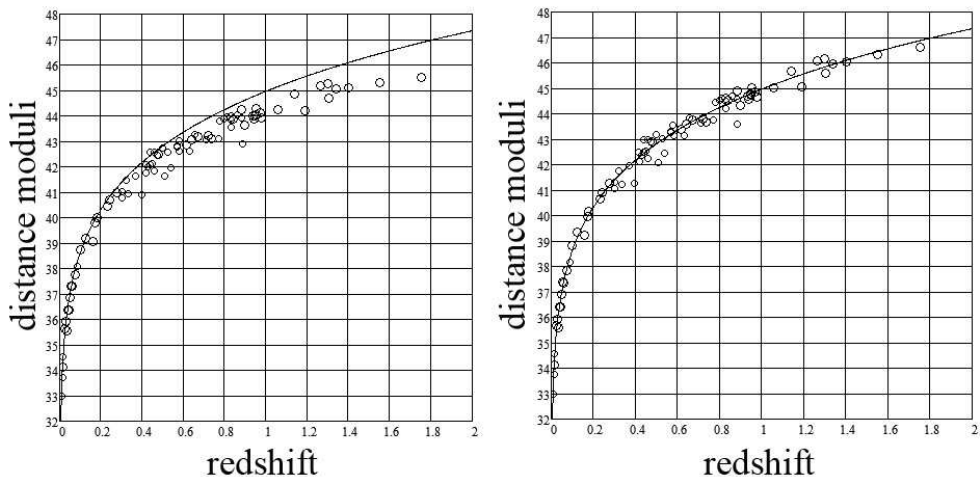}}
\caption{ The theoretical Hubble diagram $\mu_{0}(z)$ of this
model with $b=2.137$ (solid); Supernovae 1a observational data
from Table 5 of \cite{203} (circles, 82 points): corrected for
time dilation (left panel) and corrected for no time dilation
(right panel).}
\end{figure}
example on the Planck one, is not valid. Theoretical predictions
for galaxy/quasar number counts were found in this model
\cite{119} based only on the luminosity distance and the
geometrical one as functions of a redshift; there is not any
visible contradiction with observations.
\par
In the model, the luminosity distance $D_{L}$ is equal to
\cite{500}:
$$
D_{L}=a^{-1} \ln(1+z)\cdot (1+z)^{(1+b)/2} \equiv a^{-1}f_{1}(z),
$$
where $f_{1}(z)\equiv \ln(1+z)\cdot (1+z)^{(1+b)/2}$, $a=H/c,$ and
$b=2.137$ for soft radiation. Time dilation is absent in this
model; but observational data are usually corrected for this
effect of the standard cosmological model \cite{772,773}. Due to
the correction for time dilation, the observed flux is
overestimated in $(1+z)$ times, and one should correct distance
moduli $\mu_{0} = 5 \log D_{L} + 25$ as \cite{774}:
$$\mu_{0}^{'} = \mu_{0} + 2.5 \log (1+z),$$
where $\mu_{0}^{'}$ are distance moduli in any model without time
dilation. The comparison of predictions of the model with
Supernovae 1a observational data by Riess et al. \cite{203} is
shown on Figs. 1, 2. On Fig. 1, I have used the linear scale of
the vertical axis; to re-compute values of $f_{1}(z)$ from
observations, one can apply the transformation:
$$f_{1}(z)=10^{(\mu_{0}(z)- c_{1})/5},$$ where $c_{1}$ is a
constant (here its value is $c_{1}=43.4$). The left panels of
\begin{figure}[th]
\epsfxsize=9.0cm \centerline{\epsfbox{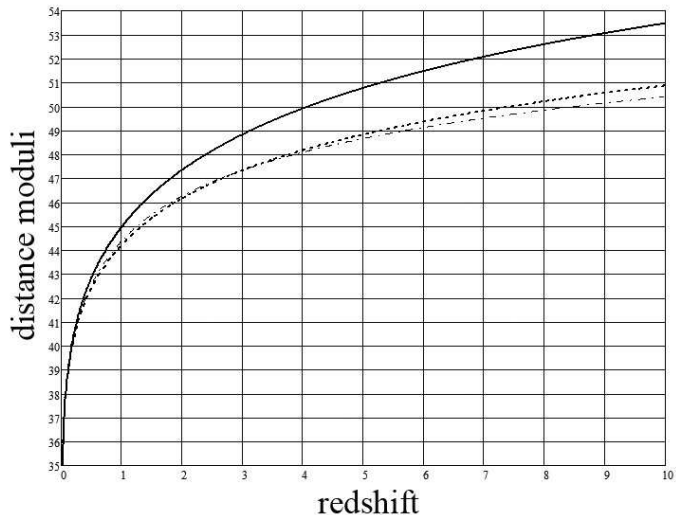}} \caption{ The
three theoretical Hubble diagrams: $\mu_{0}(z)$ of this model with
$b=2.137$ (solid); $\mu_{0}(z)$ of this model with $b=1.137$
taking into account the effect of time dilation of the standard
model (dash); $\mu_{c}(z)$ for a flat Universe with the
concordance cosmology by $\Omega_{M} = 0.27$ and $w =-1$ (dadot).}
\end{figure}
these figures are the same as Figs. 2, 3 of \cite{500}; it is
obvious now that the essential differences between predictions of
the model and observations were caused namely by the correction
for cosmological time dilation. After the correction for no time
dilation, the same observations are fitted very well with the
theoretical curve (the right panels). Some further details may be
found in my recent paper \cite{777}.
\par
As it was shown in \cite{775}, theoretical distance moduli
$\mu_{c}(z)$ for a flat Universe with the concordance cosmology by
$\Omega_{M} = 0.27$ and $w =-1$, which give the best fit to GRB
observations  by Schaefer \cite{206}, are very close to the Hubble
diagram $\mu_{0}(z)$ with $b=1.1$ of this model. From the
considered above, we see that the avoidance of the effect of
cosmological time dilation means the transition to
$b=2.137-1=1.137$ - very close to that value. We may do now some
predictions about the behavior of the universe in a frame of the
standard model for high $z$ comparing the theoretical Hubble
diagrams (see Fig. 3): $\mu_{0}(z)$ of this model with $b=1.137$
taking into account the effect of time dilation of the standard
model (dash); and $\mu_{c}(z)$ for a flat Universe with the
concordance cosmology by $\Omega_{M} = 0.27$ and $w =-1$ (dadot).
You can see a good accordance of this diagrams up to $z\approx4;$
for higher redshifts we should expect the accelerated expansion
again. The extra acceleration should decrease from big $z$ to the
smaller ones. We must bide new data from the future space missions
to verify this prediction.
\par
\begin{figure}[th]
\epsfxsize=13.0cm \centerline{\epsfbox{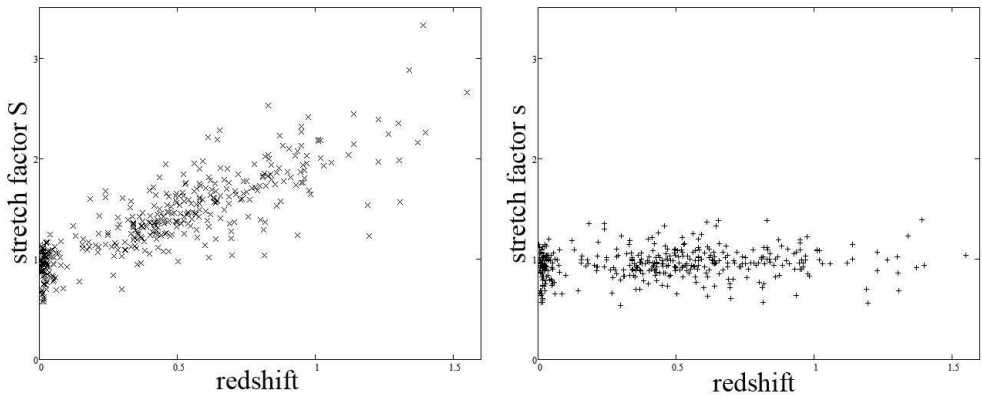}} \caption{ The
observed values of stretch factor $S$ without correction for time
dilation (left panel, $\times$) and the corresponding values of
stretch factor $s$ corrected for time dilation (right panel, $+$);
data are taken from Table 11 of \cite{776} by Kowalski et al. (the
Union compilation of SNe 1a).}
\end{figure}
Let us discuss briefly the distributions of stretch factor values
of Supernovae 1a. Supernovae light curves are characterized by
observers with the observed timescale stretch factor $S$. In the
standard cosmological model, to find the stretch factor $s$ in the
supernova rest frame one should divide $S$ by $(1+z)$, where the
factor $(1+z)^{-1}$ takes into account the effect of time dilation
\cite{772}. After it, the timescale of light curve is corrected by
the factor $S$, when its magnitude is corrected only by the
stretch factor $s$ -- in the standard model approach; but in any
model with no time dilation it is necessary to use the factor $S$
for both normalizations. On Fig. 4, the data from the Union
compilation of SNe 1a by Kowalski et al. \cite{776} are used to
show the distributions of values of $S$ and $s$ for nearby SNe 1a
(104 points with $z \leq 0.1$) and for high-z SNe 1a (294 points
with $z > 0.1$). The values of the average $\langle s \rangle$ and
$\sigma$ for $s$ are almost identical for these two subsamples:
$\langle s \rangle$ is equal to $0.91$ and $0.97$, $\sigma$ is
equal to $0.143$ and $0.144$ for nearby and remote events.
Usually, it is interpreted as the main argument in the proof that
time dilation takes place \cite{772}. But there are obvious
physical arguments to show that the distributions of the stretch
factor should be different for nearby and remote explosions: $1)$
the lower boundary of the distribution should rise with $z$ due to
increasing the luminosity distance; $2)$ the upper boundary should
rise too because we have not a possibility to observe in the local
volume very rare events, and they may be seen only in a very big
volume. We see both these expected peculiarities on the left panel
of Fig. 4, but not on the right one.

\par
In this model, energy losses of any massive body due to forehead
collisions with gravitons lead to the body acceleration by a
non-zero velocity $v$: $w_{0} = - ac^{2}(1-v^{2}/c^{2}).$  For
small velocities: $w_{0} \simeq - Hc,$ that may be connected with
the Pioneer anomaly \cite{300}.
\par Astrophysical and cosmological observations may be used not
only as confirmations of the standard model from new and new dark
sides but in another manner: to clarify and to found better our
knowledge of gravitation, perhaps, even beyond general relativity.

\end{document}